\documentclass[prl,twocolumn]{revtex4}
\usepackage{epsfig}
\usepackage{bm}
\usepackage{amsmath}

\begin{document}

\title{Eelectrokinetic turbulence and chaos in ionic migration and in microfluidic convection}

\author{A. Bershadskii}

\affiliation{
ICAR, P.O. Box 31155, Jerusalem 91000, Israel
}

\begin{abstract}

 Turbulence in the ionic migration  and in the microfluidic electro-hydrodynamic convection (mixing) under constant and low-frequency external electric field has been studied using distributed chaos approach and notion of effective (turbulent) diffusivity. Results of corresponding laboratory experiments have been also used for this purpose. It is shown that for the migration and convection (mixing) under constant and low-frequency external electric field the turbulent power spectra have stretched exponential form $E(k) \propto \exp-(k/k_{\beta})^{2/3}$. In the case of the microfluidic convection the distributed chaos is tuned to the large-scale coherent structures generated by the low-frequency oscillating external electric field ( $k_{\beta}$ is proportional to the forcing frequency).  
  
\end{abstract}

\maketitle

\section{Introduction}

  Movement of charged particles in an electrolyte solution placed in an external electric filed can be of three types: diffusion, migration and convection. The diffusion is a random movement of the particles from a region of their high concentration to regions of their lower concentration. The migration is an organized type of the movement of the charged particles due to interaction with the external electric field which does not results in a macroscopic motion of the solution, i.e. in hydrodynamic convection. The convection is an organized type of the movement of the charged particles that results in the macroscopic (hydrodynamic) motion of the solution. Both the migration and convection can become chaotic and turbulent under certain conditions. Equations describing the diffusion and (hydrodynamic) convection are well known. The hydrodynamic convection, for instance, can be described by the Navier-Stokes equations  
$$  
  \partial_t {\bf v} + ({\bf v}\cdot \nabla) {\bf v} = -\nabla p +\nu \nabla^2 {\bf v}+\mathbf{F}_e, \eqno{(1)}
$$
$$  
  \nabla \cdot \bf{v}=0   \eqno{(2)}
$$    
where $\mathbf{F}_e$ is an electrical body force. The situation with a mathematical description of migration is more complicated due to non-local character of the migration induced coupling and it is still under development (see, for instance, Ref. \cite{migr} and references therein). Anyway for the chaotic and turbulent motions both the migration and convection cases cannot be considered without a phenomenology. It will be shown in this note that the both cases can be considered with the same phenomenological notion of effective (turbulent) diffusivity \cite{my}. 

  The effective diffusivity $D$ can be estimated as
$$
D=v_cl_c   \eqno{(3)}
$$
where $v_c$ is a characteristic velocity of the migration or convection and $l_c$ is their characteristic spatial scale. 

  We will start from the migration in order to introduce naturally notion of the distributed chaos dominated by the effective diffusivity and, then, will apply this approach also to the hydrodynamical convection (mixing) in microfluidics.
  
\section{Electrochemical turbulence and distributed chaos}

  Let us start from experimental results regarding appearance of spatio-temporal chaos (defect turbulence) in an electrochemical system with non-local spatial coupling \cite{var}. In this experiment the spatio-temporal chaos was induced by the non-local migration (spatial) coupling mediated through electric potential in an electrolyte. The electric potential drop across a ring-shaped polycrystalline working electrode/electrolyte interface -  $\Phi(x,t)$, was measured by a microprobe located beneath the electrode ($x$ is the coordinate parallel to the electrode). Rotation of the ring shaped electrode over the microprobe provides the spatio-temporal evolution of $\Phi(x,t)$ and certain mass transport of ions from
the electrolyte. Two different coupling regimes - approximately local and non-local - were created by using two different distances between the working and the counter electrodes (5 and 40 mm respectively).\\

   Figure 1 shows in the semi-log scales temporally averaged Fourier (power) spectrum of the spatial profiles of the $\Phi(x,t)$ for the approximately local coupling case in the fully developed spatio-temporal chaotic regime (the spectral data were taken from Fig. 5 of the Ref \cite{var}). The dashed straight line is drawn to indicate exponential spectral decay
$$
E(k) \propto \exp-(k/k_c)   \eqno{(4)}
$$
where $k$ is wavenumber and $k_c = 1/l_c$ (position of the $k_c$ in the figure is indicated by the dotted arrow). 

  The exponential frequency spectra are well known characteristic of the temporal deterministic chaos of dynamical 
systems \cite{fm}-\cite{b}. The spatial (wavenumber) exponential spectra for the spatio-temporal chaos near onset of isotropic homogeneous turbulence were observed recently in direct numerical simulations \cite{kds}.

   When we are going from the local coupling to the non-local one the characteristic scale $k_c=1/l_c$ becomes fluctuating and in order to calculate the spatial spectrum one should use an ensemble average  
$$
E(k) \propto \int_0^{\infty} P(k_c) \exp -(k/k_c)dk_c  \propto \exp-(k/k_{\beta})^{\beta}  \eqno{(5)}
$$
where $P(k_c)$ is the ensemble distribution of the fluctuating characteristic scale $k_c =1/l_c$ and the stretched exponential in the right-hand side of the equation is a natural generalization of the exponential spectrum.\\
\begin{figure} \vspace{-1.5cm}\centering
\epsfig{width=.45\textwidth,file=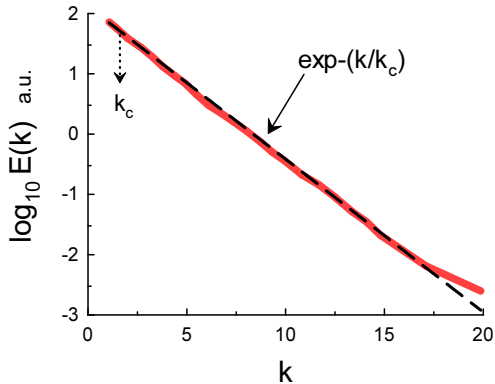} \vspace{-4cm}
\caption{Temporally averaged Fourier (power) spectrum of the spatial profiles of the $\Phi(x,t)$ for the approximately local coupling case.} 
\end{figure}
\begin{figure} \vspace{-0.4cm}\centering
\epsfig{width=.45 \textwidth,file=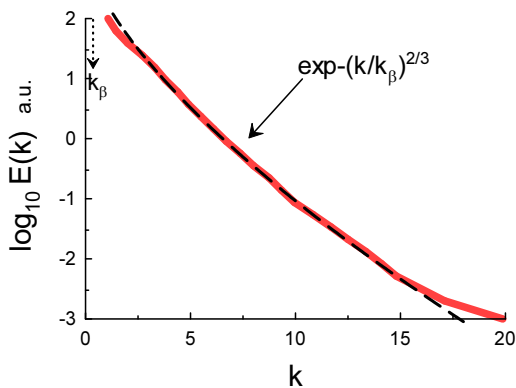} \vspace{-4cm}
\caption{As in Fig. 1 but for non-local situation.} 
\end{figure}

 Asymptote of the distribution $P(k_c)$ at $k_c \rightarrow \infty$ can be found from the Eq. (5)
$$
P(k_c) \propto k_c^{-1 + \beta/[2(1-\beta)]}~\exp(-bk_c^{\beta/(1-\beta)}) \eqno{(6)}
$$
with $b$ as a constant \cite{jon}, whereas it follows from from Eq. (3) that for constant diffusivity $D$ the characteristic wavenumber $k_c=1/l_c$ has the same distribution as the characteristic velocity $v_c$ - Gaussian, for instance. The asymptotic distribution Eq. (6) is Gaussian if $\beta = 2/3$, i.e.
$$
E(k) \propto \exp-(k/k_{\beta})^{2/3}  \eqno{(7)}
$$
in this case. 

   Figure 2 shows in the semi-log scales temporally averaged Fourier (power) spectrum of the spatial profiles of the $\Phi(x,t)$ for the non-local coupling case in the fully developed spatio-temporal chaotic regime (the spectral data were taken from Fig. 5 of the Ref \cite{var}). The dashed straight line is drawn to indicate the stretched exponential spectrum Eq. (7) and the dotted arrow indicates position of the scale $k_{\beta}$.

\section{Electrokinetic turbulence in microfluidic convection (mixing)}

  At certain conditions the organized motion of the ions interacting with the external electric filed results in macroscopic motion of the solution - hydrodynamic convection, which can be described by the Navier-Stokes Eqs. (1-2). An interesting and practically significant case of such motion is fast mixing of two fluids with different conductivities in an external electric field. For protein folding with fast kinetics, for instance, the enhanced mixing is highly demanded. It is also important for lab-on-a-chip (see, for instance, Ref. \cite{cy} and references therein). On the other hand, Reynolds number in the microchannels is usually small (of order one), that demands certain specific efforts in order to enhance the mixing (it is interesting that the spectrum Eq. (7) was considered for the first time for viscosity dominated distributed chaos \cite{b2}). The pressure drive of the microchannel flow alone is insufficient to produce a fast (turbulent) mixing, but addition of the electrical body force $\mathbf{F}_e$ in the Eq. (1) can solve the problem. In this case \cite{wyz}
$$ 
\bf{F}_e = (\rho_f/\rho) {\bf E}  \eqno{(8)}
$$
where ${\bf E}$ is the external electric field, $\rho$ is fluid density, the initial free charge density in the solution
$$
\rho_f = \varepsilon {\bf E} \cdot\nabla \sigma/\sigma,   \eqno{(9)}
$$
$\varepsilon$ is the permittivity and $\sigma$ is the electric conductivity of the electrolyte, respectively.

   Corresponding electric Rayleigh number 
$$
Ra_e = \varepsilon E_0 d^2 (\sigma_2-\sigma_1)/ \sigma_1 \rho \nu D_e   \eqno{(10)}
$$
where $E_0$ is the root-mean square value of the nominal electric field strength , $d$ is the width of the
channel at the entrance (see Fig. 3 adapted from the Ref. \cite{wyz}), $D_e$ is an effective diffusivity, $\sigma_1$ and $\sigma_2$ are the conductivities of the two mixing streams. At sufficiently high $Ra_e$ complex three-dimensional multiscale flow can be developed even at small Reynolds numbers.\\

   Figure 4 shows kinetic energy spectrum measured (with the nanoscopic Laser
Induced Fluorescence Photobleaching Anemometer \cite{ku}-\cite{kqw}), in the the microfluidic chip's experiment described in the Ref. \cite{wyz} (the spectral data were taken from Fig. 6 of the Ref. \cite{wyz}, see also Fig. 3 adapted from the Ref. \cite{wyz} with a sketch of setup of the experimental microfluidic chip). 
\begin{figure} \vspace{+1cm}\centering
\epsfig{width=.7\textwidth,file=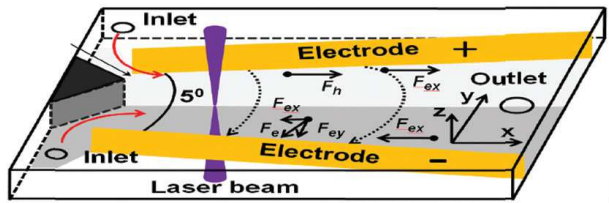} \vspace{-14.07cm}
\caption{Sketch of setup of the experimental microfluidic chip.} 
\end{figure}
\begin{figure} \vspace{-0.5cm}\centering
\epsfig{width=.44 \textwidth,file=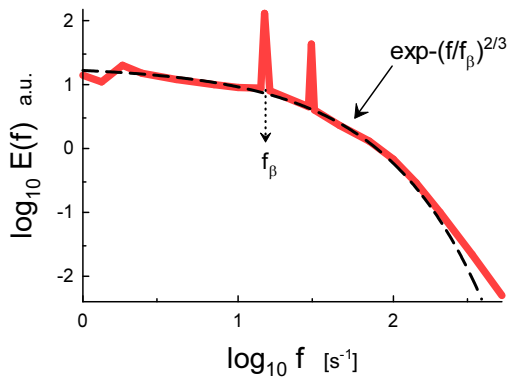} \vspace{-4cm}
\caption{Kinetic energy spectrum in the microfluidic electrokinetic convection (mixing) at $Ra \simeq 2.8 \times 10^6$.} 
\end{figure}

   The well known Taylor hypothesis relates the frequency $f$ to the wavenumber $k$ \cite{my}: 
$$
k=\frac{2\pi f}{U}   \eqno{(11)}
$$
where $U$ is the mean velocity, and it can be used for comparison with the Eq. (7). The dashed curve in the Fig. 4 is drawn for this purpose. The dotted arrow indicates position of the $k_{\beta} = 2\pi f_{\beta}/U$, where the $f_{\beta}=15 Hz$ 
is the forcing frequency of the external electric field in this case (corresponding to the main peak in the power spectrum in Fig. 4). One can see that the distributed chaos is tuned to the forcing frequency (to the large-scale coherent structures produced by the forcing oscillation, cf. also Ref. \cite{b}). For very high forcing frequency of $10^5$ Hz, i.e. without the forcing of the large-scale coherent structures, the spectrum is substantially different (see Fig. 4 in the Ref. \cite{wyz}).\\

   In present case the electric Rayleigh number $Ra_e \simeq 2.8 \times 10^6$ and it is interesting to compare spectrum shown in the Fig. 4 with kinetic energy spectrum in the fully developed thermal (Rayleigh-B\'{e}nard) convection at thermal Rayleigh number $Ra = 10^8$ obtained in direct numerical simulation described in Ref. \cite{vvs}. This spectrum is shown in Fig. 5 and the dashed curve is drawn to indicate correspondence to the Eq. (7). 
\begin{figure} \vspace{-1.5cm}\centering
\epsfig{width=.42 \textwidth,file=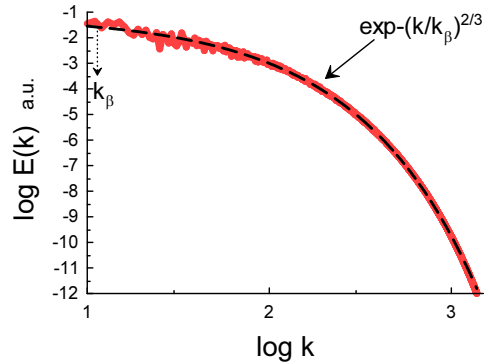} \vspace{-4cm}
\caption{Kinetic energy spectrum in the thermal convection at $Ra=10^8$.} 
\end{figure}

\section{Acknowledgement}

I thank R. Samuel and M.K. Verma for sharing their data.

\end{document}